# On the Vineyard Formula for the Pre-Exponential Factor in the Arrhenius Law

M.M. Maslov*, L.A. Openov, and A.I. Podlivaev

*National Research Nuclear University "MEPhI", Kashirskoe sh. 31, Moscow, 115409 Russia*
* e-mail: Mike.Maslov@gmail.com

ABSTRACT

By the example of several typical thermally activated processes in atomic clusters, organic molecules, and nanostructures, it is shown that calculations of the corresponding pre-exponential factors in the Arrhenius law according to the Vineyard formula are in good agreement with the molecular dynamics simulation data for temperature dependences of characteristic times of these processes. This "static" approach (together with the determination of the activation energy through the examination of the potential energy hypersurface) provides information on kinetic characteristics of the system without resorting to numerical simulation of the time evolution, which requires large computer resources.

## 1. INTRODUCTION

The temperature dependence of the rate $k$ of one or another thermally activated process in solids or atomic clusters (diffusion, decomposition, isomerization, etc. [1–4]) is often described by the Arrhenius law

$$k(T) = A \exp\left(-\frac{E_a}{k_\text{B} T}\right), \qquad (1)$$

where $T$ is the temperature, $k_\text{B}$ is the Boltzmann constant, $E_a$ is the activation energy (the minimum energy required to be transferred to the system to overcome an energy barrier on the corresponding reaction path), and $A$ is the frequency factor with dimension of s$^{-1}$. In computer simulation, the dependence $k(T)$ can be found using the molecular dynamics method and directly determining the characteristic time of the process $\tau = k^{-1}$ at different temperatures [2, 5–9]. For example, $\tau$ is the time of the complete and irreversible loss of the initial structure upon decomposition of the cluster [2, 5, 6] or the time of transition to the isomer with a different atomic configuration upon isomerization [7, 9]. In both cases, $\tau$ is actually the lifetime of either the cluster or its isomer. As an illustration, Fig. 1 presents the data obtained from the "computer experiment" for the lifetime of a nonclassical fullerene $C_{46}$ [9]. It can be seen from this figure that, in a rather wide temperature range, the dependence of the logarithm of the lifetime $\tau$ on the inverse temperature is well approximated by a straight line in accordance with the Arrhenius law (1). The activation energy can be determined from the slope of this line, and the frequency factor can be found from the point of its intersection with the ordinate axis (in this case, $E_a = 3.1 \pm 0.2$ eV and $A = 10^{16.0 \pm 0.3}$ s$^{-1}$, respectively [9]). However, it should be noted that the "dynamic" approach to the simulation of thermally activated processes has a significant drawback: it requires very large computer resources. For this reason, for example, the *ab initio* molecular dynamics allows examination of the evolution of a system consisting of ~100 atoms only for a very short time of ~1 ps [10], which is not sufficient to collect the necessary statistics and to determine the temperature dependence of the lifetime τ. Even with the use of simplified interatomic interaction models [11, 12], the molecular dynamics simulation requires very long calculations [2, 5, 6, 9] (especially at low temperatures, when the value of τ is exponentially large; see formula (1)). Hence, the problem of determining the dependence τ(T) becomes extremely labor-consuming (and often even unfeasible) from a purely technical point of view.

An alternative approach is the static simulation: the determination of the values of $E_a$ and $A$ in formula (1) by analyzing the dependence of the potential energy of the system $E_{pot}$ on the coordinates $\mathbf{R}_i$ of the constituent atoms in the vicinity of the local minimum of $E_{pot}$ (corresponding to the initial configuration) and the nearest local maximum, which corresponds to the saddle point of the hypersurface $E_{pot}(\{\mathbf{R}_i\})$. In this case, the activation energy $E_a$ is assumed to be equal to the height of the corresponding energy barrier, and the frequency factor $A$ is calculated according to the Vineyard formula [13], which for an $N$-atom cluster with zero values of the total momentum and the total angular momentum has the form

$$A = \frac{\prod_{i=1}^{3N-6} \nu_i}{\prod_{i=1}^{3N-7} \nu'_i}, \qquad (2)$$

where $\nu_i$ are the eigenfrequencies of vibrations of the cluster in the state corresponding to the minimum of the potential energy $E_{pot}$ for all $3N - 6$ normal coordinates, and $\nu'_i$ are the frequencies of vibrations at the saddle point corresponding to the maximum of $E_{pot}$ for one normal coordinate and to the minimum for all the other coordinates (since one of the $3N - 6$ frequencies $\nu'_i$ is imaginary, it is not included in the denominator of expression (2) for $A$). In the simulation of a macroscopic sample by an $N$-atom supercell with periodic boundary conditions, the numbers of nonzero real frequencies $\nu_i$ and $\nu'_i$ in expression (2) are equal to $3N - 3$ and $3N - 4$, respectively [14].

Since, in the static simulation, it is much easier to determine the activation energy (barrier height) than the frequency factor, the latter parameter is often not calculated but is simply assumed to be equal (or close) to the characteristic frequency $f$ of atomic vibrations; i.e., it is taken to be $A \approx 10^{13}$ s$^{-1}$ [15]. If we follow this logic, the value of $A$ obviously cannot exceed a certain maximum frequency, which for most of solids and clusters with covalent interatomic bonds is $f_{max} \approx 3 \times 10^{13}$ s$^{-1}$. One of our recent works [9] was originally prepared for publication in other journal, however it was rejected by one of the referees who believed that our results might be wrong since the value of the frequency factor $A \sim 10^{18}$ s$^{-1}$, which we obtained, appeared to be unrealistically large for the reviewer.

However, in the literature, there are many examples of thermally activated processes for which $A \gg f_{max}$. For example, in the diffusion of adatoms on the Cu(100) surface, the value of $A$ is larger than $10^{15}$ s$^{-1}$ [3], whereas for the thermal fragmentation of the fullerene $C_{60}$ both theory and experiment give $A \approx 10^{20}$ s$^{-1}$ [5, 16–18]. In the latter case, of course, one could argue that the fragmentation of the cluster is not an elementary process and occurs in several stages. Nonetheless, it seems quite obvious that, for a many-particle physical system, the frequency factor according to formula (2) is not reduced to a particular frequency, but is determined by a specific form of the function $E_{pot}(\{\mathbf{R}_i\})$ of a large number of variables, not only in a local minimum but also in a local maximum of this function. From formula (2), in particular, it follows that the large frequency factor $A$ can be explained by the presence of nearly planar regions on the potential energy hypersurface $E_{\text{pot}}(\{\mathbf{R}_i\})$ in the vicinity of the saddle point. As a result, one or several frequencies $v'_i$ in the denominator of formula (2) are very small.

The question arises as to whether the frequency factor $A$ calculated from formula (2) agrees with the frequency factor found by means of the direct molecular dynamics determination of the temperature dependence of the lifetime. The purpose of this work is to perform a comparative analysis of the data obtained from the static and dynamic simulations of several typical thermally activated processes. As will be shown below, within the error of "computer experiment" both approaches lead to almost identical results.

## 2. COMPUTATIONAL TECHNIQUES

In this work, we will restrict our consideration to the case of carbon and hydrocarbon systems. For these systems, there have been developed the effective tight-binding (orthogonal [11] and nonorthogonal [12]) methods, which rank below the *ab initio* methods in accuracy but are much less demanding of computer resources and, therefore, in the static simulation, make it possible to investigate in detail the function $E_{\text{pot}}(\{\mathbf{R}_i\})$ for a system consisting of 100–1000 atoms and, in the dynamic simulation, to examine the evolution of a system consisting of 10–100 atoms for a sufficiently long (on the atomic scale) time ranging from 10 ns to 1 μs. For structural and energy characteristics of the clusters (including fullerenes), molecules, and macroscopic objects (graphene, graphane), these methods give values in good agreement with first-principles calculations and experimental data [12].

In the dynamic simulation, we determined the frequency factor $A$ by analyzing the temperature dependence of the lifetime τ, which was found using the molecular dynamics method, as in the example presented above (Fig. 1). We used the dependences τ(T) obtained earlier for several typical thermally activated processes [6–9]. In each case, the time step was less than 1 fs, and the results were presented in the form of a computer animation. This allowed us, first, with a good accuracy, to determine the lifetime at different temperatures and, second, to find the atomic configurations corresponding to saddle points of $E_{\text{pot}}$ (for more details, see [9, 19]). In the static simulation, the frequency factor was calculated according to formula (2). For this purpose, we determined the frequencies $v_i$ and $v'_i$ by the diagonalization of the Hessian, i.e., the matrix of the second derivatives of $E_{\text{pot}}(\{\mathbf{R}_i\})$, with respect to the coordinates of atoms at the minimum and saddle point, respectively (the configurations used in this case were found in the dynamic simulation). The zero frequencies corresponding to rotational and/or translational degrees of freedom were disregarded.

## 3. RESULTS AND DISCUSSION

We compared the data obtained from the dynamic and static simulations of various thermally activated processes occurring in the cubane $C_8H_8$, nonclassical fullerene $C_{46}$, fullerene $C_{60}$, and graphone. The results obtained for each of these systems are presented below.

### 3.1. Cubane $C_8H_8$

In the cubane $C_8H_8$ [20], carbon atoms are located at vertices of the cube, so that the C–C covalent bonds form an angle of 90°, unusual for carbon systems, which is energetically unfavorable.

The hydrogen atoms located on the main diagonals of the cube stabilize the atomic configuration (Fig. 2) corresponding to not the global minimum but to a local minimum of the potential energy as a function of the atomic coordinates. Although the cubane is a metastable cluster, it is sufficiently stable, i.e., retains the structure at temperatures well above room temperature and can even form a molecular crystal, namely, the solid cubane s-$C_8H_8$ with a melting temperature of approximately 400 K [21]. This stability of the cubane is explained by a high activation energy of cubane decomposition $E_a = 1.9 \pm 0.1$ eV [6].

The cubane decomposition often starts with a thermally activated transition to the isomer *syn*-tricyclooctadiene, which rapidly (for 0.1–1.0 ps) transforms either into cyclooctatetraene or into bicyclooctatriene [6]. In the simulation of the time evolution of a heated cubane, we have never observed transitions of these isomers back to the cubane; i.e., the isomerization of the cubane is actually equivalent to its decomposition (final products in this case are usually molecules of benzene and acetylene). The frequency factor determined from the temperature dependence of the lifetime of the cubane to the onset of isomerization/decomposition is $A = 10^{16.0 \pm 0.4}$ s$^{-1}$ [6].

Before comparing this value of $A$ with the calculations from formula (2), we note that this formula determines the frequency factor in the rate (1) of a thermally activated process for only one of the possible channels. Even if we restrict our consideration to the case of the most probable transition paths (characterized by the lowest activation energy), we still need to take into account that there can be several such paths due to the symmetry of the system. Since the isomerization/decomposition of cubane $C_8H_8$ occurs upon breaking any of the twelve equivalent C–C bonds, the frequency factor in the total rate of isomerization/decomposition (it is this rate that is determined in the dynamic simulation) should be 12 times higher than that calculated from formula (2) for one channel. Consequently, the result of the calculation from formula (2) must be multiplied by the "degeneracy factor" $g = 12$. As a result, we obtain $A = 10^{16.48}$ s$^{-1}$, which is in good agreement with the molecular dynamics simulation data.

### 3.2. Nonclassical Fullerene $C_{46}$

In the nonclassical fullerene $C_{46}$, the C–C covalent bonds form not only pentagons and hexagons, as in conventional fullerenes, but also a square (Fig. 3). The theory has predicted [22] that this isomer is thermodynamically more stable than its classical analogue and is an exception to the general rule of energy instability of nonclassical fullerenes. According to the molecular dynamics data [9], the decomposition of this cluster begins with the breaking of one of the C–C bonds forming a square. This leads first to the formation of an octagonal "window" and then to a rapid transition (for a time of ~1 ps) to the classical isomer. Since subsequently the initial atomic configuration is no longer recovered, the time of isomerization of the nonclassical fullerene $C_{46}$ is actually its lifetime τ (the cluster can long retain a spheroidal shape; i.e., the isomerization does not lead to the decomposition, as in the above case of the cubane $C_8H_8$).

The analysis of the temperature dependence of the lifetime τ, which was obtained from the dynamic simulation (Fig. 1), demonstrated that the frequency factor for the isomerization of the nonclassical fullerene $C_{46}$ is $A = 10^{16.0 \pm 0.3}$ s$^{-1}$ [9]. The calculations performed using formula (2) lead to the frequency factor $A = 10^{16.13}$ s$^{-1}$, where it is taken into account that the "isomerization factor" $g$ in this case is equal to four, which is the number of C–C bonds that can be broken during the isomerization. Thus, there is excellent agreement between the results obtained in the static approach and the molecular dynamics data.

### 3.3. Fullerene $C_{60}$

In the fullerene $C_{60}$, the C–C bonds form twenty hexagons and twelve pentagons isolated from each other [23]. During heating of this cluster [5], the defect formation begins with the Stone–Wales transformation [24], which consists in rotating one of the C–C bonds common for hexagons through an angle of 90°. This leads to the formation of an isomer with two pairs of adjacent pentagons (Fig. 4), and the energy increases by ~1.4 eV. This isomer is long-lived and can even transform back into the perfect fullerene (the $C_{60}$ fullerene loses a spheroidal shape only after the accumulation of a sufficiently large amount of Stone–Wales defects along with defects of other types [5]). The height of the energy barrier for the Stone–Wales transformation is approximately equal to 6.5 eV [25]. The saddle point of $E_{pot}(\{\mathbf{R}_i\})$ for this barrier corresponds to the rotation of the C–C bond through an angle of 45°.

From the analysis of the temperature dependence of the time of isomerization of the fullerene $C_{60}$ through this channel, which was obtained by the molecular dynamics method, it follows that the frequency factor of the isomerization process is $A = 10^{17.3 \pm 0.6}$ $s^{-1}$ [7]. The calculation performed using formula (2) with the degeneracy factor $g = 60$ (during the Stone–Wales transformation, each of the thirty C–C bonds common for hexagons can be rotated in two directions) leads to the frequency factor $A = 10^{17.06}$ $s^{-1}$, which is in complete agreement with the dynamic simulation data.

### 3.4. Graphone

Graphone [26] is a graphene [27] on one side of which hydrogen atoms are adsorbed on every second carbon atom (Fig. 5). According to the theory [26], graphone is a magnetic semiconductor with a band gap $E_g \approx 0.5$ eV. Practical use of graphone is hampered by its thermal instability due to the very low activation energy of hopping of hydrogen atoms between the neighboring carbon atoms: $E_a = 0.05 \pm 0.01$ eV [8]. The migration of hydrogen atoms leads to a distortion of the graphone structure and to an uncontrollable change in its magnetic characteristics.

In the coordinate space, the saddle points of $E_{pot}(\{\mathbf{R}_i\})$, which determine the energy barriers $U = 0.058$ eV for the hydrogen migration, are located in approximately the middle between the carbon atoms, i.e., between the local minima of $E_{pot}(\{\mathbf{R}_i\})$. The calculation performed using formula (2) for the migration process gives the frequency factor $A = 10^{13.7}$ $s^{-1}$, which almost completely coincides with the value $A = 10^{13.5 \pm 0.1}$ $s^{-1}$ obtained by the dynamic simulation [8].

All the above results are summarized in the table. As can be seen, there is a very good quantitative (within the statistical error) agreement between the static and dynamic simulation data for the frequency factor $A$ in the Arrhenius law (1). It is worth noting that this agreement holds over a very wide range of $A$ values (four orders of magnitude); i.e., the Vineyard formula (2) is applicable to the description of both the "slow" and "fast" processes.

It can also be seen from the table that, for all the considered thermally activated processes, the activation energy $E_a$ (determined in common with the frequency factor from the temperature dependence of the characteristic time of the process) is consistent with the corresponding energy barrier height $U$, which is calculated using the static simulation. This usually takes place when the process occurs predominantly through one channel. In the case of several channels, the rate of the process is determined by the sum of terms of the type (1).

Note that we observed good agreement between the dynamic and static simulation data in the study of the thermal stability of other systems as well, for example, fullerene $C_{20}$ [28], nitrogen cubane $N_8$ [29], methylcubane $C_9H_{10}$ [30], etc.

## 4. CONCLUSIONS

The main result obtained in this work is the demonstration of the fact that the frequency factors of typical thermally activated processes, which were calculated by the Vineyard formula, are in good agreement with direct numerical simulation of the time evolution of the system by the molecular dynamics method. Since the activation energy involved in the Arrhenius law for the rate of the process is usually close to the height of the corresponding energy barrier, this activation energy and the frequency factor can be determined by analyzing the potential energy hypersurface. Thus, the static approach makes it possible to determine temperature dependences of the characteristic times of various thermally activated processes without resorting to long-term simulation of the dynamics of the system in real time, which requires large computer resources (and, sometimes, is unfeasible).

The disadvantage of the static approach is the lack of a priori information about the paths of the system evolution during heating and, hence, about the atomic configurations corresponding to saddle points of the potential energy. In some cases, the configuration can be determined from symmetry considerations; however, this cannot always be done, especially if the transition process (decomposition, isomerization, etc.) includes a large group of atoms. The solution of the problem is to combine the dynamic and static simulations: first, to investigate the dynamics of the system at high temperatures and to determine the main transition paths and the atomic configurations encountered in these paths and, then, to analyze the potential energy hypersurface in the vicinity of the obtained metastable and saddle configurations, to find the energy barrier heights, and to calculate the corresponding frequency factors.

The problem is significantly complicated if the thermally activated process can proceed through a large number of different channels, which takes place, for example, in the isomerization of clusters with low symmetry. In this case, one should either restrict himself to analyzing only a part of the channels (at the expense of the accuracy) or use only the dynamic simulation (with large expenditure of the computer time). When choosing the optimal strategy, it is necessary to take into account both the specifics of the physical system and the features of the studied process.

## ACKNOWLEDGMENTS

This study was supported by the Russian Foundation for Basic Research (project no. 12-02-00561).

# REFERENCES

1. C. Wert and C. Zener, Phys. Rev. **76**, 1169 (1949).
2. M. M. Maslov, Russ. J. Phys. Chem. B **4**, 170 (2010).
3. F. Montalenti and A. F. Voter, Phys. Status Solidi B **226**, 21 (2001).
4. A. I. Podlivaev and L. A. Openov, Phys. Solid State **54**, 1507 (2012).
5. L. A. Openov and A. I. Podlivaev, JETP Lett. **84**, 68 (2006).
6. M. M. Maslov, D. A. Lobanov, A. I. Podlivaev, and L. A. Openov, Phys. Solid State **51**, 645 (2009).
7. A. I. Podlivaev and K. P. Katin, JETP Lett. **92**, 52 (2010).
8. A. I. Podlivaev and L. A. Openov, Semiconductors **45**, 958 (2011).
9. L. A. Openov, A. I. Podlivaev, and M. M. Maslov, Phys. Lett. A **376**, 3146 (2012).
10. X.-L. Sheng, H.-J. Cui, F. Ye, Q.-B. Yan, Q.-R. Zheng, and G. Su, J. Appl. Phys. **112**, 074315 (2012).
11. C. H. Xu, C. Z. Wang, C. T. Chan, and K. M. Ho, J. Phys.: Condens. Matter **4**, 6047 (1992).
12. M. M. Maslov, A. I. Podlivaev, and L. A. Openov, Phys. Lett. A **373**, 1653 (2009).
13. G. H. Vineyard, J. Phys. Chem. Solids **3**, 121 (1957).
14. A. I. Podlivaev and L. A. Openov, Phys. Solid State **55**, 2592 (2013).
15. A. Santana, A. M. Popov, and E. Bichoutskaia, Chem. Phys. Lett. **557**, 80 (2013).
16. C. Lifshitz, Int. J. Mass Spectrom. **198**, 1 (2000).
17. S. Tomita, J. U. Andersen, K. Hansen, and P. Hvelplund, Chem. Phys. Lett. **382**, 120 (2003).
18. K. Hansen, E. E. B. Campbell, and O. Echt, Int. J. Mass Spectrom. **252**, 79 (2006).
19. M. M. Maslov, A. I. Podlivaev, and L. A. Openov, Phys. Solid State **53**, 2532 (2011).
20. P. E. Eaton and T. W. Cole, Jr., J. Am. Chem. Soc. **86**, 962 (1964).
21. M. A. White, R. E. Wasylishen, P. E. Eaton, Y. Xiong, K. Pramod, and N. Nodari, J. Phys. Chem. **96**, 421 (1992).
22. J. An, L.-H. Gan, X. Fan, and F. Pan, Chem. Phys. Lett. **511**, 351 (2011).
23. H. W. Kroto, J. R. Heath, S. C. O'Brien, R. F. Curl, and R. E. Smalley, Nature **318**, 162 (1985).
24. A. J. Stone and D. J. Wales, Chem. Phys. Lett. **128**, 501 (1986).
25. A. I. Podlivaev and L. A. Openov, JETP Lett. **81**, 533 (2005).
26. J. Zhou, Q. Wang, Q. Sun, X. S. Chen, Y. Kawazoe, and P. Jena, Nano Lett. **9**, 3867 (2009).
27. K. S. Novoselov, A. K. Geim, S. V. Morozov, D. Jiang, Y. Zhang, S. V. Dubonos, I. V. Grigorieva, and A. A. Firsov, Science **306**, 666 (2004).
28. I. V. Davydov, A. I. Podlivaev, and L. A. Openov, Phys. Solid State **47**, 778 (2005).
29. L. A. Openov, D. A. Lobanov, and A. I. Podlivaev, Phys. Solid State **52**, 201 (2010).
30. M. M. Maslov, Russ. J. Phys. Chem. B **3**, 211 (2009).

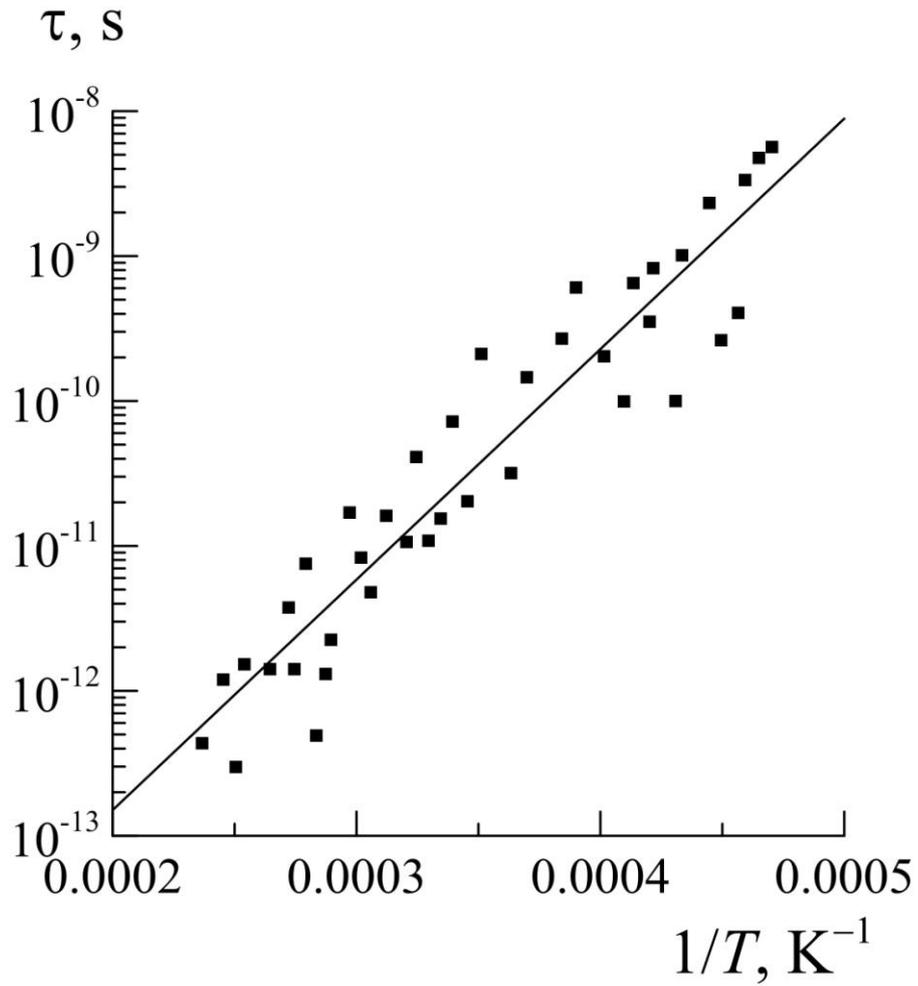

**Fig. 1.** Dependence of the lifetime $\tau$ of the nonclassical fullerene $C_{46}$ until the onset of isomerization on the inverse temperature $T^{-1}$ (plotted on a logarithmic scale). Points are the results of the computer simulation, and the solid line is the linear approximation by the least-squares method.

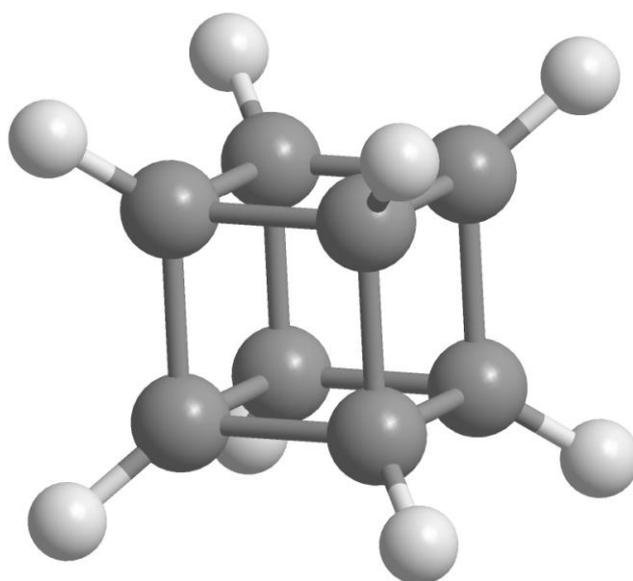

**Fig. 2.** Cubane $C_8H_8$. Gray spheres are carbon atoms, and white spheres are hydrogen atoms.

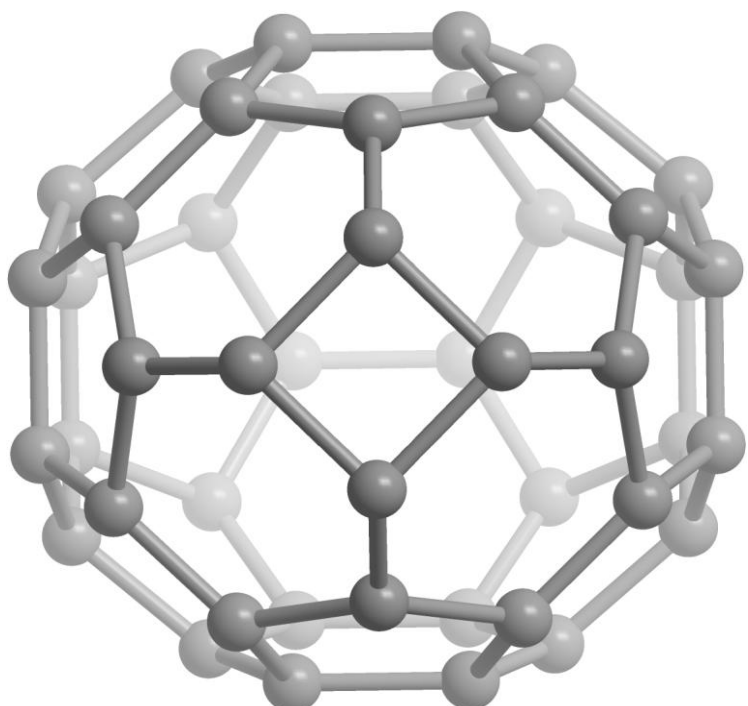

**Fig. 3.** Nonclassical fullerene $C_{46}$.

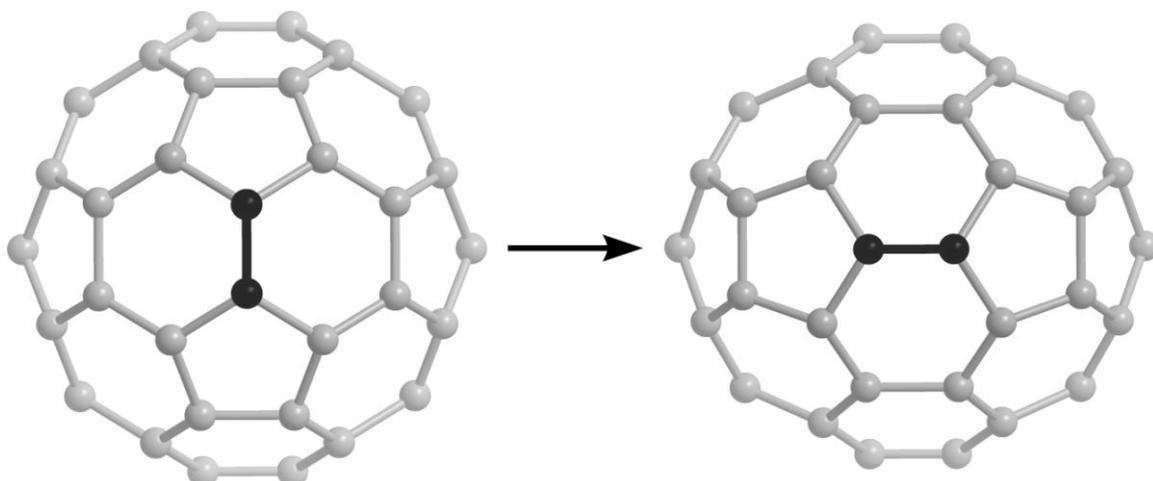

**Fig. 4.** Stone–Wales transformation in fullerene $C_{60}$. The black color indicates the C–C bond, which rotates through an angle of 90° during the transformation. For clarity, atoms and bonds on the distant plan are not shown.

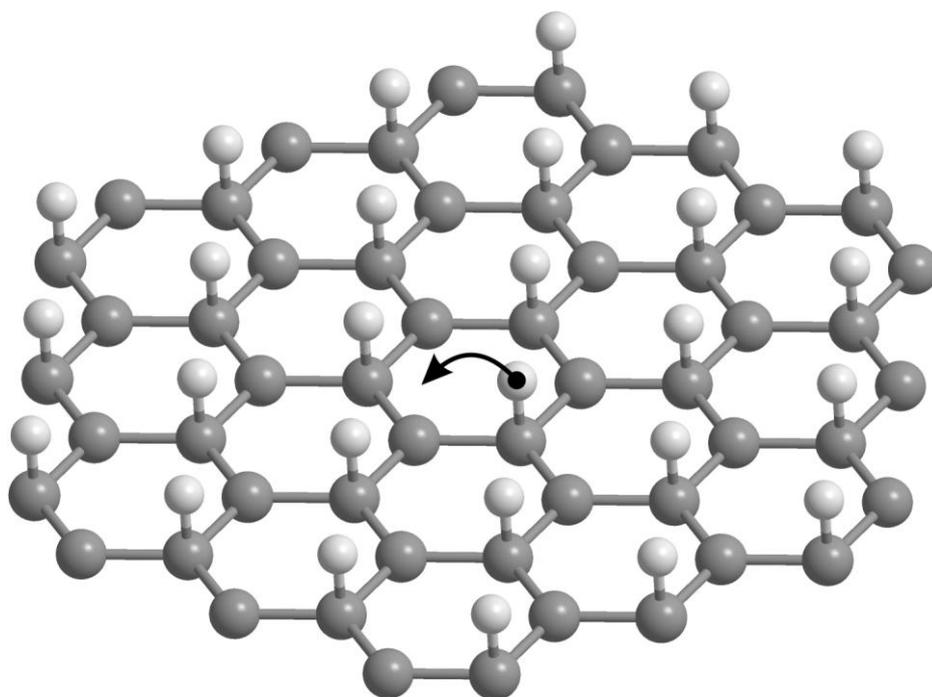

**Fig. 5.** Fragment of the graphone. Gray spheres are carbon atoms, and white spheres are hydrogen atoms. The arrow indicates the direction of a possible hopping (migration) of one of the hydrogen atoms.

TABLE

| Physical object | Process | Activation energy $E_a$, eV (DS) | Barrier height $U$, eV (SS) | Frequency factor $A$, s$^{-1}$ | |
| --- | --- | --- | --- | --- | --- |
| | | | | DS | SS |
| Cubane $C_8H_8$ (Fig. 2) | Isomerization/ decomposition | $1.9 \pm 0.1$ | 1.6 | $10^{16.0 \pm 0.4}$ | $10^{16.48}$ |
| Nonclassical fullerene $C_{46}$ (Fig. 3) | Isomerization | $3.1 \pm 0.2$ | 2.9 | $10^{16.0 \pm 0.3}$ | $10^{16.13}$ |
| Fullerene $C_{60}$ (Fig. 4) | Isomerization | $6.3 \pm 0.4$ | 6.5 | $10^{17.3 \pm 0.6}$ | $10^{17.06}$ |
| Graphone (Fig. 5) | Migration | $0.05 \pm 0.01$ | 0.058 | $10^{13.5 \pm 0.1}$ | $10^{13.7}$ |

Designations: DS is dynamic simulation (direct determination of the temperature dependence of the lifetime, molecular dynamics method), and SS is static simulation (analysis of the potential energy hypersurface, Vineyard formula).